\documentstyle[preprint,aps]{revtex}
\renewcommand{\baselinestretch}{1}
\def\ra{\rangle}
\def\la{\langle}
\def\ver{\arrowvert}

\begin{document}
\title{Uncertainty relation of Anandan-Aharonov and Intelligent states}       
\author{Arun Kumar Pati}
\address{SEECS, Dean Street, University of Wales, Bangor LL 57 1UT, UK.}
\address{and}
\address{Theoretical Physics Division, 5th Floor, Central Complex}
\address{BARC, Mumbai-400 085, India}
               
\maketitle

\begin{abstract}
The quantum states which satisfy the equality in the
generalised uncertainty relation are called intelligent states.
We prove the existence of intelligent states for the
Anandan-Aharonov uncertainty relation based on the geometry of
the quantum state space
for arbitrary parametric evolutions of quantum
states when the initial and final states are non-orthogonal.
\end{abstract}

\vskip .5cm

email:akpati@sees.bangor.ac.uk
\vskip 1cm

%\begin{multicols}{2}

   In recent years the study of geometry of the quantum state space and its implications
have gained much importance. The introduction of Riemannian metric structure by
Provost  and Valle \cite{pv} and Fubini-Study metric  by Anandan and Aharonov \cite{aa,an}
into the projective Hilbert space of the quantum system has attracted a lots of
attention.
The relation between geometric distance function and geometric phase was studied and
the equivalence of the above two metric strctures (up to a scale factor) was
pointed out \cite{ap}.
The introduction of the length of the curve \cite{ap,akp} has provided us a new
way of  understanding geometric phases
in quantum systems. Subsequently, the metric structures were
generalised to mixed states by Anandan \cite{ana} and the statistical
distinguishability was used to define a metric structre by Braunstein and Caves
\cite{sam}.  Later, the Fubini-Study metric was generalised to non-unitary and
non-linear quantum systems, and a metric approach to generalised geometric phase
was proposed \cite{akp2}.
One of the outcome of the geometric approach is
the parameter-based uncertainty relation (PBUR) in quantum theory. This is
often useful when we do not have a Hermitian operator canonical conjugate
to another operator which represent a physical quantity of our interest. The
vivid example is the quest for time-energy  uncertanity relation, when we do
not have a Hermitian time operator canonical conjugate to energy.

  In this letter we study the intelligent states (to be defined soon) for
the Aharonov-Anandan uncertainty relation and prove the existence of such states
when the initial and final states  are non-orthogonal during an arbitrary
parametric evolution of a quantum system.

   To briefly recall the essential geometric ideas, let us consider a quantum
system $\cal S$ whose state vector
$\ver \psi(t) \ra \in {\cal H} = {\cal C}^N $ evolves in time from time $t_1$ to
$t_2$. Geometrically the state is represented by a point in the projective
Hilbert space ${\cal P} = {\cal H} - \{ 0 \} / {\cal C}^*$, where ${\cal C}^*$
is a group of non-zero complex numbers. The time evolution of the system
gives us a curve $C$ in $\cal H$, i.e. $C: t \rightarrow \ver \psi(t) \ra,
t_1 \le t \le t_2$.  Since  $\cal H$ is Riemannian this curve has a length
which has been extensively studied by the present author \cite{ap,akp,akp1}. The
Hilbert space curve can be projected onto a curve ${\hat C} = \Pi(C)$ via a
projection map $\Pi: \cal H \rightarrow \cal P$. The projected curve $\hat C$
has a length which is induced from the inner product of the Hilbert space and
is given by the Fubini-Study metric \cite{pv,aa,an,ap,akp,ana,sam,akp2,akp1}

\begin{equation}
S= {2 \over \hbar} \int_{t_1}^{t_2} \Delta H(t) dt,
\end{equation}
where $\Delta H(t) =  \bigg[ \la \psi(t) \ver H(t)^2 \ver \psi(t) \ra
-   \la \psi(t) \ver H(t) \ver \psi(t) \ra^2 \bigg]^{1 \over 2}$ is the usual uncertainty
in the energy of the system. This distance is independent of a particular
Hamiltonian used to evolve the quantum system and is invariant under a gauge
transformation. On the other hand if $S_0$ is the shortest distance between
the initial and final states joining the points $\Pi(\ver \psi(t_1) \ra)$ and
$\Pi(\ver \psi(t_2) \ra)$ then the distance measured by Fubini-Study metric
is alway greater than the geodesic distance connecting the initial and
final points, where $S_0$ is given by the Bargmann angle \cite{akp} formula
$\cos^2 {S_0 \over 2}=  \ver  \la \psi(t_1) \ver \psi(t_2) \ra \ver ^2$. This gives
$S \ge S_0$ and the equality holds only for those states that evolve along
a geodesic in $\cal P$. The above geometric reasoning is the basis of
the Anandan-Aharonov uncertainty relation for time and energy,
which is given by

\begin{equation}
\la \Delta H(t) \ra \Delta t \ge { h \over 4},
\end{equation}
where $ \la \Delta H(t) \ra = {1 \over (t_2 - t_1)} \int_{t_1}^{t_2} \Delta H(t) dt$
is the time average of the energy uncertainty and
$\Delta t = {\pi \over S_0} (t_2 - t_1)$ is the ``uncertainty in time''. When
the initial and final states are orthogonal (which are distinguishable by
quantum mechanical tests) then the shortest distance is $\pi$. In this case
time-energy uncertainty relation takes a simple form (for a time-independent
Hamiltonian) as

\begin{equation}
\Delta H \Delta t \ge { h \over 4},
\end{equation}
where $\Delta t = (t_2 -t_1)$ is the ordinary time difference that is required
to make a transition to a orthogonal state.
There are various aproaches to time-energy uncertainty relation and to
estimate the time required to make a transition
to orthogonal states in the literatue \cite{lv,jdf,vb,ml,js}.
These estimations are not only of theoretical interest but have important
implication on how fast one can do a computation in a quantum computer \cite{ml,sl}.
The advantage of the geometric
uncertainty relation is that we do not have to look for a Hermitian operator
for time. It can remain just as a parameter and uncertainty in the parameter
would mean how good we can estimate it given a certain amount of uncertanity
in the conjugate variable. This fact can be grounded by the observation that
we can go beyond the time-energy uncertainty relation. If we have any
continuous parameter $\lambda$ and any Hermitian observable $A(\lambda)$ which
is the generator of the parametric evolution, then a similar geometric
reasoning would give us

\begin{equation}
\la \Delta A(\lambda) \ra \Delta \lambda \ge { h \over 4},
\end{equation}
where $ \la \Delta A(\lambda) \ra = {1 \over (\lambda_2 - \lambda_1)}
\int_{\lambda_1}^{\lambda_2} \Delta A(\lambda) d \lambda$
is the parameter average of the observable uncertainty and
$\Delta \lambda = {\pi \over S_0} (\lambda_2 - \lambda_1) $ is the
scaled displacement in the space of the conjugate variable of $A$. This
generalised uncertainty relation would hold for position-momentum, phase-number
or any possible combinations. Recently, Yu \cite{yu} has discussed the PBUR
for position-momentum case. We have also proposed a geometric uncertainty relation
without reference to any Hermitian operator and shown that (2) and (4)
follows as a special case of our generalised uncertainty relation \cite{akp3}.

   It is known that given two non-commuting observables we can derive an
uncertainty relation for them and the class of states that satisfy the equality
sign in the inequality are called {\it intelligent states} \cite{ag,acs}.
So the natural question would be what are the intelligent states for geometric
uncertainty relation? To characterise these states in general is quite
difficult. But one pretty observation is that since the initial and final
state can be connected by a geodesic, states which satisfy the equality
have to be equivalent (i.e. belong to a ray) to a state that satisfies
the geodesic equation. Since
solution to geodesic equation lies in a two-dimensional (real)
subspace \cite{akp1,ms},
the intelligent states should also belong to a two-dimensional   subspace
${\cal H}_s$. This will be clear as we proceed to our next discussion.
Recently, Horesh and Mann \cite{hm} have constructed a state that
satisfies the equality in PBUR when the initial and final states are orthogonal.
These state indeed belong to a 2-dim. subspace of the full Hilbert space.
However, their construction fails when the initial and final states are
non-orthogonal. In this letter we find the intelligent states for PBUR or
geometric uncertainty relation (GUR) when the initial and final states
are non-orthogonal. We prove a theorem
conecerning the intelligent states for both orthogonal and non-orthogonal cases.
This will provide an answer to the question what kind of states do evolve along
a geodesic in the projective Hilbert space.

%Geodesics are special paths in
%the projective Hilbert space along which a quantum state preserves its phase.
%So from a quantum information processing point of view (where a state can
%undergo phase shift error) keeping track of
%geodeics are important, because this will ensure not to acquire any phase.

   Let us consider a quantum system represented by a state
$\ver \psi(\lambda) \ra$. Let $A$ be the Hermitian generator of the
parameter translation.
For simplicity we assume that $A$ does not depend on $\lambda$. The unitary
evolution operator $U= exp(-iA {\lambda \over \hbar})$ generates the evolution
path in $\cal H$ starting from a given state $\ver \psi(0) \ra$. The
length of the actual evolution path in $\cal P$ is measured by
Fubini-Study metric. This can be defined via length \cite{akp1}
of a parallel-transported
vector $\ver \bar{\psi}(\lambda) \ra$, where  $\ver \bar{\psi}(\lambda) \ra
= exp({i \over \hbar} \int_0^{\lambda} \la \psi(\lambda') \ver A \ver
 \psi(\lambda') \ra d \lambda') \ver \psi(\lambda) \ra$. This satisfies the
 parallel
 transport condition $\la \bar{\psi}(\lambda) \ver {d \over d\lambda}
\bar{\psi}(\lambda) \ra = 0$, which physically means the vector
$\ver \bar{\psi}(\lambda) \ra$ does not rotate locally but undergoes a net
rotation when comes back to its original point in the projective Hilbert
space.
The geodesic in $\cal P$ can be defined as the one for
which length of the path traced by the parallel-transported vector is minimum.
The length of the curve traced by the parallel-transported state is given by
\cite{akp,akp1}

\begin{equation}
 l(\bar{\psi}) = \int_0^{\lambda} ~ \la {d \over d \lambda'} \bar{\psi}(\lambda')
\ver {d \over d \lambda'} \bar{\psi}(\lambda') \ra^{\frac{1}{2} } ~d\lambda'.
\end{equation}

The variational calculation gives an equation for geodesic (see for details
\cite{akp1} and \cite{ms})

\begin{equation}
{d^2 \ver \bar{\psi}(\lambda) \ra \over d\lambda^2} + v^2 \ver \bar{\psi}(\lambda) \ra = 0,
\end{equation}
where $v = {\Delta A \over \hbar}$. The geodesic equation has a general solution

\begin{equation}
\ver \bar{\psi}(\lambda) \ra = \cos (v \lambda) \ver \bar{\psi}(0) \ra +
{1 \over v} \sin (v \lambda) \ver \dot{\bar{\psi}}(0) \ra,
\end{equation}
where $ \la \bar{\psi}(0) \ver \bar{\psi}(0) \ra = 1$,
$ \la \dot{\bar{\psi}}(0) \ver \dot{\bar{\psi}}(0) \ra = v^2$, $v$ being the
speed of the system point in ${\cal P}$ and $\ver \dot{\bar{\psi}}(0) \ra$
 denotes derivative wrt the parameter.
Though the geodesic on ${\cal P}$ is obtained by projecting a curve ${\bar C}:
\lambda \rightarrow \ver \bar{\psi}(\lambda) \ra$ in
${\cal H}$, we have expressed the equation of geodesic (6) in terms of the
state $\ver \bar{\psi}(\lambda) \ra$. This is because, given a projected
path $\Pi(C_g)$ in ${\cal P}$
there is a horizontal lift of the path in ${\cal H}$ such that the length of
the path ${\bar C}$ is same as that of the $\Pi(C_g)$.

Thus the states
which satisfy geodesic equation are linear superposition of two mutually
orthogonal vectors. The corresponding one-dimensional projector $\rho(\lambda)
=\ver \bar{\psi}(\lambda) \ra \la \bar{\psi}(\lambda) \ver$
belongs to a two-dimensional real space.
Note that  $\rho(\lambda)$ can be expressed as

\begin{equation}
\rho(\lambda) = \cos^2 (v \lambda) \rho_{11} +
\sin^2 (v \lambda) \rho_{22} + \cos (v \lambda) \sin (v \lambda)
 (\rho_{12} + \rho_{21} ),
\end{equation}
where $\rho_{ij} = \ver \bar{\psi}_{i} \ra \la \bar{\psi}_{j} \ver$,
$\ver \bar{\psi}_{i} \ra = \ver \bar{\psi}(0) \ra $ and  $\ver \bar{\psi}_{j} \ra
= {1 \over v} \ver \dot{\bar{\psi}}(0) \ra$ ($ij=1,2$). This is  a two-dimensional geometric
quantity.
This is consistent with the well known
fact \cite{mc} that the geodesics
on the unit sphere are intersections of two-dimensional 
subspace ${\cal H}_s=
span \{ \ver \bar{\psi}_{i} \ra ,\ver \bar{\psi}_{j} \ra \}$ 
with the sphere, i.e. they live in a two-dimensional manifold. Now we state the
following theorem.\\

Theorem: {\it If $\ver \psi(\lambda) \ra $ is the intelligent state 
which satisfy the equality in PBUR and $\ver \bar{\psi}(\lambda) \ra$ is the
parallel transported state obtained from  $\ver \psi(\lambda) \ra $  then
$\ver \bar{\psi}(\lambda) \ra$ satisfy
the equation of geodesic and can always be expressed in the form (7)
for both the orthogonal and non-orthogonal initial and final states.}\\

     First we prove it for the case when initial and final states are
orthogonal. From the work of Horesh and Mann \cite{hm} we know that all
states of the form

\begin{equation}
\ver \psi(\lambda) \ra = {1 \over \sqrt 2} \biggl( exp(-{i \over \hbar} a_i \lambda )\ver \psi_i \ra +
 exp(-{i \over \hbar} a_j \lambda )\ver \psi_j \ra \biggr), ~~~~~~i \not= j
\end{equation}
are the {\it only} intelligent states which satisfy the equality in (4). It is assumed that $A$ has a complete
basis of normalised state $\{\ver \psi_i \ra \}_{i \in I}$ with non-degenerate
spectrum $\{a_i\}_{i \in I}$, with $I$ a set of quantum numbers.
This state at parameter value $\lambda = {\pi \hbar \over (a_j - a_i)}$
becomes orthogonal to the initial state. The parallel transported state
$\ver \bar{\psi}(\lambda) \ra = e^{{i \over \hbar} \la A \ra \lambda} \ver \psi(\lambda) \ra$
is given by

\begin{equation}
\ver \bar{\psi}(\lambda) \ra = {1 \over \sqrt 2}( exp(-{i \over 2\hbar}(a_i - a_j) \lambda )\ver \psi_i \ra +
exp({i \over 2\hbar}(a_i - a_j) \lambda ) \ver \psi_j \ra ), ~~~~~~i \not= j.
\end{equation}
It is easy to check that this state satisfies geodesic equation (6). Further, by
noting that $v = \Delta A = {1 \over 2}(a_i - a_j)$ we can reexpress
the above state as
$\ver \bar{\psi}(\lambda) \ra = 
\cos (v \lambda) \ver \bar{\psi}(0) \ra +
{1 \over v} \sin (v \lambda) \ver \dot{\bar{\psi}}(0) \ra $, where
$\ver \bar{\psi}(0) \ra = {1 \over \sqrt 2}( \ver \psi_i \ra +  \ver \psi_j \ra)$
and ${1 \over v} \ver \dot{\bar{\psi}}(0) \ra = 
{i \over \sqrt 2}(\ver \psi_i \ra -  \ver \psi_j \ra)$.
This completes the proof when the initial and final states are orthogonal.\\

  However, the states given in (9) {\em do not satisfy the equality }
sign in (4) when the initial and
final states are {\em non-orthogonal}. In quantum theory
the study of non-orthogonal states
are crucial because they display variety of non-classical features. Moreover,
non-orthogonal states cannot be distinguished by ordinary measurement
processes without error. Therefore, it is important to know what kind of states
can evolve along a geodesic connecting two non-orthogonal states.
Now we seek those states which satisfy
the equality in PBUR or GUR of Anandan-Aharonov for non-orthogonal
initial and final states. In general for arbitrary
generator $A$ it is not possible to find a state which will be an
intelligent one. So, the question is what condition should we impose on the
Hermitian operator $A$, such that we will be able to find intelligent states
for non-orthogonal initial and final states. The problem is more acute when
we do not know the complete set of eigenstates of the operator $A$. But suppose
we know the spectrum of a part of the operator $A$. Then
we propose the following:

Proposition: {\it If the Hermitian generator $A$ of the parametric evolution
can be split
into two parts $A_0 + A_1$ such that $A_0$ has a complete basis of normalised
eigenstates
$\{\ver \psi_i \ra \}_{i \in I}$ with degenerate
spectrum $\{a_0\}$, with $I$ a set of quantum numbers and $A_1$ has matrix
elements $(A_1)_{ii} = 0 = (A_1)_{jj}$ and $(A_1)_{ij} = (A_1)_{ji} = a_1$, then
all states of the form

\begin{eqnarray}
\ver \psi(\lambda) \ra = e^{{-i \over \hbar} a \lambda}  (\cos(a_1 {\lambda \over \hbar})
\ver \psi_i \ra  - i \sin(a_1 {\lambda \over \hbar})
\ver \psi_j \ra ), ~~~~~~i \not= j \nonumber\\
\end{eqnarray}
are intelligent states for non-orthogonal initial and final states.}

To prove this, we focus on the parametric evolution of the state $\ver \psi(
\lambda) \ra$ in the two-dimensional subspace spanned by vectors $\{
 \ver \psi_i \ra, \ver \psi_j \ra \}$. Since $\lambda$ is a continuous parameter
we can solve the equation of motion for $\ver \psi(\lambda) \ra$, i.e. 
$i \hbar {d \over d \lambda} \ver \psi(\lambda) \ra = A \ver \psi(\lambda) \ra$
with a given initial condition. Without loss of generality we assume that
initially the state is in $\ver \psi_i \ra$. The state at any other parameter
value can be written as  $\ver \psi(\lambda) = c_i(\lambda) \ver \psi_i \ra +
c_j(\lambda) \ver \psi_j \ra$. Then the solution to evolution equation gives

\begin{eqnarray}
c_i(\lambda) = e^{-i a \lambda \over \hbar} \cos(a_1 {\lambda \over \hbar}), \nonumber\\
c_j(\lambda) = -i e^{-i a \lambda \over \hbar} \sin(a_1 {\lambda \over \hbar}).
\end{eqnarray}
With the help of these amplitudes we find the uncertainty in the operator
$A$. It is given by

\begin{equation}
{\Delta A}^2 = a_0^2 + a_1^2 + 4 a_0 a_1 Re(c_i^* c_j) - (a_0 + 2 a_1 Re(c_i^* c_j))^2
  = a_1^2 ,
\end{equation}
where we have used the fact that $c_i^* c_j$ is purely imaginary quantity.
Next we calculate the shortest distance along the geodesic using
$S_0 = 2 \cos^{-1}( \ver  \la \psi(\lambda_1) \ver \psi(\lambda_2) \ra \ver)$,
which is ${2 \over \hbar} a_1 \lambda_2$. We have taken initial parameter value
$\lambda_1 = 0$. Therefore, the ``uncertainty in the parameter'' is
$\Delta \lambda = {\pi \over S_0} \lambda_2 = {\pi \over 2 a_1} \hbar$. Thus
the lhs of the PBUR of Anandan-Aharonov becomes

\begin{equation}
\Delta A \Delta \lambda  = a_1.{\pi \over 2 a_1} \hbar = {h \over 4}.
\end{equation}
This proves that the states (11) in our proposition are indeed intelligent states
for arbitrary parametric evolution of quantum systems, when the initial and
final points are non-orthogonal.

  Now we prove our theorem for non-orthogonal cases. It is easy to see that
the parallel-transported state in this case is given by 
$\ver \bar{\psi}(\lambda) \ra =  (\cos(a_1 {\lambda \over \hbar})
\ver \psi_i \ra  - i \sin(a_1 {\lambda \over \hbar})
\ver \psi_j \ra )$. This satisfies the geodesic equation (6). Also, the above state
can be expressed as
$\ver \bar{\psi}(\lambda) \ra = 
\cos (v \lambda) \ver \bar{\psi}(0) \ra +
{1 \over v} \sin (v \lambda) \ver \dot{\bar{\psi}}(0) \ra $, where one can check that
$\ver \bar{\psi}(0) \ra = \ver \psi_i \ra $
and ${1 \over v} \ver \dot{\bar{\psi}}(0) \ra = 
-i \ver \psi_j \ra$. This ends the proof.

    We remark that any other state which is
intelligent has to be equivalent (where an equivalence relation $\sim$ means
$\ver \psi(\lambda) \ra \sim \ver \psi(\lambda)' \ra$ if
$\ver \psi(\lambda)' \ra = c \ver \psi(\lambda) \ra$, where $c$ is a complex
number of unit modulus, i.e. $c \in U(1)$ group) to the state given in (11). Any state of the
form $\ver \psi \ra = c_i \ver \psi_i \ra  + c_j \ver \psi_j \ra  +
c_k \ver \psi_k \ra, (i \not= j \not= k)$ will  not be an intellgent state
because it will not satisfy geodesic equation. 

      To conclude this letter, we have proved a theorem which says that
in general when the intelligent
states satisfy geometric uncertainty relation, the corresponding
parallel transported states satisfy geodesic
equation. This guarantees that intelligent states belong to a two-dimensional
manifold. For arbitrary parametric evolution of a quantum system we have
found the explicit form of intelligent states by imposing certain condition
on the generator when the initial and final states are {\it non-orthogonal}.  The
intelligent states for orthogonal case found in \cite{hm} are distinct
from that of the non-orthogonal case found here.

\vskip 1cm

{\bf Acknowledgements:} I am grateful to S. L. Braunstein for his general support and
ESPRC for financial support. I also thank the referee for his suggestions to
improve the clarity of the paper.

\renewcommand{\baselinestretch}{1}

%\end{multicols}

\newpage

\end{document}